# Coupling-induced nonunitary and unitary scattering in anti-$\mathcal{PT}$-symmetric non-Hermitian systems

H. S. Xu and L. Jin [*]

*School of Physics, Nankai University, Tianjin 300071, China*

We investigate the properties of two anti-parity-time (anti-$\mathcal{PT}$)-symmetric four-site scattering centers. The anti-$\mathcal{PT}$-symmetric scattering center may have imaginary couplings, real couplings, and real on-site potentials. The only difference between the two scattering centers is the coupling between two central sites of the scattering center, which plays a crucial role in determining the parity of anti-$\mathcal{PT}$ symmetry and significantly affects the scattering properties. For the imaginary coupling, the even-parity anti-$\mathcal{PT}$-symmetric scattering center possesses nonunitary scattering and the difference between the reflection and transmission is unity; for the real coupling, the odd-parity anti-$\mathcal{PT}$-symmetric scattering center possesses unitary scattering and the sum of the reflection and transmission is unity. The coupling-induced different scattering behaviors are verified in the numerical simulations. Our findings reveal that a significant difference in the dynamics can be caused by a slight difference between two similar anti-$\mathcal{PT}$-symmetric non-Hermitian scattering centers.

## I. INTRODUCTION

In quantum mechanics, the Hermitian Hamiltonian of a quantum system ensures the real energy spectrum and the conservation of a probability current. In addition, non-Hermitian parity-time ($\mathcal{PT}$), symmetric Hamiltonians can possess real spectra [1–6] and have been experimentally realized in many different physical systems [7–13]. The non-Hermitian Hamiltonians possess peculiar features that have no counterpart in the Hermitian Hamiltonians [14–19]. The non-Hermitian systems have been widely investigated and many interesting properties have been revealed in the past two decades. The nonorthogonal eigenstate and nonunitary time evolution [20–23], eigenstate coalescence and exceptional point (EP) enhanced sensing [24–28], and the exotic topology of EPs have been predicted and successfully demonstrated in many experimental platforms [29–35]; particularly, realization in the quantum regime [36,37]. These have boosted the development of the complex extension of quantum mechanics and greatly deepen our understanding of non-Hermitian systems—research interest in non-Hermitian systems continuously increases in the material, engineering, and physics science.

Quantum transport and wave propagation in non-Hermitian systems exhibit intriguing scattering behaviors; for example, coherent perfect absorption [38–41], unidirectional invisibility [12,42], unidirectional reflectionless [43], and unidirectional lasing at the spectral singularity [44,45]. In a non-Hermitian scattering center, scattering is usually nonunitary and the non-Hermiticity can lead to different reflection and/or transmission for the wave injected in opposite directions. It is worth mentioning that $\mathcal{PT}$ symmetry plays an important role. The reflection-$\mathcal{PT}$ symmetry protects the transmission to be identical and the axial-$\mathcal{PT}$ symmetry protects the reflection to be identical for the wave injected in opposite directions [46].

The scattering properties of a system with $\mathcal{PT}$ symmetry are explicit but anti-$\mathcal{PT}$ symmetry as a counterpart of $\mathcal{PT}$ symmetry has rarely been investigated [47–57]. Recently, imaginary coupling has been demonstrated in atomic vapors [48], electrical circuits [58], and optical waveguides [59]. In the coupled resonator array, the linking resonator with dissipation can be adiabatically eliminated to realize the imaginary coupling between two adjacent resonators [56,60]. Anti-$\mathcal{PT}$ symmetry ensures the real part of the energy spectrum either being zero or opposite in pairs; in contrast, $\mathcal{PT}$ symmetry ensures the imaginary part of the energy spectrum either being zero or opposite in pairs. The dissipation induces nontrivial topology in anti-$\mathcal{PT}$-symmetric systems [56,57], which greatly differs from the unaffected topology in $\mathcal{PT}$-symmetric systems [61].

In this paper, we investigate the properties of two anti-$\mathcal{PT}$-symmetric four-site scattering centers. The non-Hermitian scattering centers are reciprocal and their Hamiltonians are transpose invariant. The two scattering centers possess different parities of anti-$\mathcal{PT}$ symmetry. The parities are fully determined by the central coupling between the central two sites of the scattering centers. If the central coupling is imaginary, the four-site scattering center has even-parity anti-$\mathcal{PT}$-symmetry, the scattering is nonunitary, and the reflection ($R$) and transmission ($T$) satisfy $R - T = 1$. If the central coupling is real, the four-site scattering center has odd-parity anti-$\mathcal{PT}$-symmetry, the scattering is unitary, and the reflection and transmission satisfy $R + T = 1$ even though the scattering center is non-Hermitian.

The remainder of this paper is organized as follows. In Sec. II, we present the even-parity anti-$\mathcal{PT}$-symmetric four-site scattering center with all the couplings being imaginary and we demonstrate the nonunitary scattering dynamics. In Sec. III, we demonstrate the unitary scattering in the odd-parity anti-$\mathcal{PT}$-symmetric four-site scattering center by replacing the imaginary coupling between the two central sites with the real coupling. In Sec. IV, we analyze the influence of the coupling between the central two sites, which determines the parity of the anti-$\mathcal{PT}$ symmetry. We perform numerical simulation to elucidate our findings and exhibit the dynamic features of nonunitary scattering at the spectral singularity. In Sec. V, we discuss the experimental implementation of the proposed anti-$\mathcal{PT}$-symmetric systems in coupled resonator optical waveguides (CROWs). Finally, we summarize the results and conclude in Sec. VI.

---

[*]jinliang@nankai.edu.cn

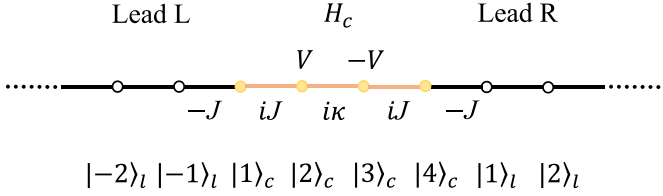

FIG. 1. Schematic of the anti-$\mathcal{PT}$-symmetric system with nonunitary scattering. The orange part is the scattering center. The scattering center is connected to two semi-infinite tight-binding chains. The non-Hermiticity arises from the reciprocal imaginary couplings (orange lines) between the sites of scattering center.

## II. NONUNITARY SCATTERING

The discrete lattice model can characterize the continuum systems with periodical potential in the Wannier representation under the tight-binding approximation [62]. In this section, we study the properties of an anti-$\mathcal{PT}$-symmetric non-Hermitian scattering center, which is a four-site scattering center with imaginary coupling. The imaginary coupling can be realized through dissipation in the non-Hermitian systems [56]. The schematic of the anti-$\mathcal{PT}$-symmetric non-Hermitian system is shown in Fig. 1. The Hamiltonian reads

$$H = H_{\text{lead}} + H_{\text{in}} + H_{\text{c}}. \quad (1)$$

$H_{\text{lead}}$ indicates the input and output leads, which contains two semi-infinite chains with uniform coupling strength $J$ in the form of

$$H_{\text{lead}} = -J \sum_{j=1}^{\infty} (|j\rangle_1\langle j+1|_1 + |-j\rangle_1\langle -j-1|_1 + \text{H.c.}), \quad (2)$$

where $|j\rangle_1$ is the basis of the lead site and represents the single excitation subspace. The connection Hamiltonian is

$$H_{\text{in}} = -J(|-1\rangle_1\langle 1|_{\text{c}} + |4\rangle_{\text{c}}\langle 1|_1 + \text{H.c.}), \quad (3)$$

where $|1\rangle_{\text{c}}$ and $|4\rangle_{\text{c}}$ are the sites of the scattering center $H_{\text{c}}$ that connected to the input and output leads.

The scattering center is embedded in a uniform one-dimensional chain and consists of four sites, the Hamiltonian reads

$$H_{\text{c}} = iJ(|1\rangle_{\text{c}}\langle 2|_{\text{c}} + |2\rangle_{\text{c}}\langle 1|_{\text{c}} + |3\rangle_{\text{c}}\langle 4|_{\text{c}} + |4\rangle_{\text{c}}\langle 3|_{\text{c}})$$
$$+ i\kappa(|2\rangle_{\text{c}}\langle 3|_{\text{c}} + |3\rangle_{\text{c}}\langle 2|_{\text{c}}) + V(|2\rangle_{\text{c}}\langle 2|_{\text{c}} - |3\rangle_{\text{c}}\langle 3|_{\text{c}}), \quad (4)$$

where $V$ is the real on-site potential and $\kappa$ is coupling strength, and $|j\rangle_{\text{c}}$ ($j = 1, 2, 3, 4$) is the basis of the center site. The matrix form of $H_{\text{c}}$ is

$$H_{\text{c}} = \begin{pmatrix} 0 & iJ & 0 & 0 \\ iJ & V & i\kappa & 0 \\ 0 & i\kappa & -V & iJ \\ 0 & 0 & iJ & 0 \end{pmatrix}. \quad (5)$$

The parity operator $\mathcal{P}$ is defined as the operation of spatial inversion, in this model

$$\mathcal{P} = \begin{pmatrix} 0 & 0 & 0 & 1 \\ 0 & 0 & 1 & 0 \\ 0 & 1 & 0 & 0 \\ 1 & 0 & 0 & 0 \end{pmatrix}. \quad (6)$$

The time-reversal operator $\mathcal{T}$ is the complex conjugation operation defined as $\mathcal{T}i\mathcal{T}^{-1} = -i$. Under these definitions, the scattering center $H_{\text{c}}$ possesses anti-$\mathcal{PT}$-symmetry, which satisfies $(\mathcal{PT})H_{\text{c}}(\mathcal{PT})^{-1} = -H_{\text{c}}$.

Now we calculate the reflection and transmission coefficients of $H_{\text{c}}$ for the left and right inputs, respectively. We suppose that the wave function for the left input is denoted as $\psi_{\text{L}}^k$ and for the right input is denoted as $\psi_{\text{R}}^k$ for the lead site $|j\rangle_1$, where $k$ is the dimensionless wave vector for the input and the incoming plane wave will be reflected and transmitted by the scattering center. The wave functions are in the form of

$$\psi_{\text{L}}^k(j) = \begin{cases} e^{ik(j-1)} + r_{\text{L}} e^{-ik(j-1)}, & j < 0 \\ t_{\text{L}} e^{ik(j+1)}, & j > 0, \end{cases} \quad (7)$$

$$\psi_{\text{R}}^k(j) = \begin{cases} t_{\text{R}} e^{-ik(j-1)}, & j < 0 \\ e^{-ik(j+1)} + r_{\text{R}} e^{ik(j+1)}, & j > 0, \end{cases} \quad (8)$$

where $r_{\text{L}}$ ($t_{\text{L}}$) and $r_{\text{R}}$ ($t_{\text{R}}$) are the reflection (transmission) coefficients for the left and right inputs, respectively. The lead of the model is a uniformly coupled tight-binding chain, so we obtain the dispersion relation $E = -2J \cos k$ from the Schrödinger equations for the lead Hamiltonian $H_{\text{lead}}$.

Therefore, the Schrödinger equations for the scattering center $H_{\text{c}}$ are

$$-J\psi_{\text{L(R)}}^k(-1) + iJ\psi_{\text{c}}^k(2) = E\psi_{\text{c}}^k(1), \quad (9)$$

$$iJ\psi_{\text{c}}^k(1) + V\psi_{\text{c}}^k(2) + i\kappa\psi_{\text{c}}^k(3) = E\psi_{\text{c}}^k(2), \quad (10)$$

$$i\kappa\psi_{\text{c}}^k(2) - V\psi_{\text{c}}^k(3) + iJ\psi_{\text{c}}^k(4) = E\psi_{\text{c}}^k(3), \quad (11)$$

$$iJ\psi_{\text{c}}^k(3) - J\psi_{\text{L(R)}}^k(1) = E\psi_{\text{c}}^k(4). \quad (12)$$

For the left input, we set the wave functions as $\psi_{\text{L}}^k(-1) = e^{-2ik} + r_{\text{L}} e^{2ik}$ and $\psi_{\text{L}}^k(1) = t_{\text{L}} e^{2ik}$ in Eq. (7); we can obtain $\psi_{\text{c}}^k(1) = e^{-ik} + r_{\text{L}} e^{ik}$ and $\psi_{\text{c}}^k(4) = t_{\text{L}} e^{ik}$ from the Schrödinger equations in the sites $|-1\rangle_1$ and $|1\rangle_1$; for the right input, as we set the wave functions $\psi_{\text{R}}^k(-1) = t_{\text{R}} e^{2ik}$ and $\psi_{\text{R}}^k(1) = e^{-2ik} + r_{\text{R}} e^{2ik}$ in Eq. (8), we can obtain $\psi_{\text{c}}^k(1) = t_{\text{R}} e^{ik}$ and

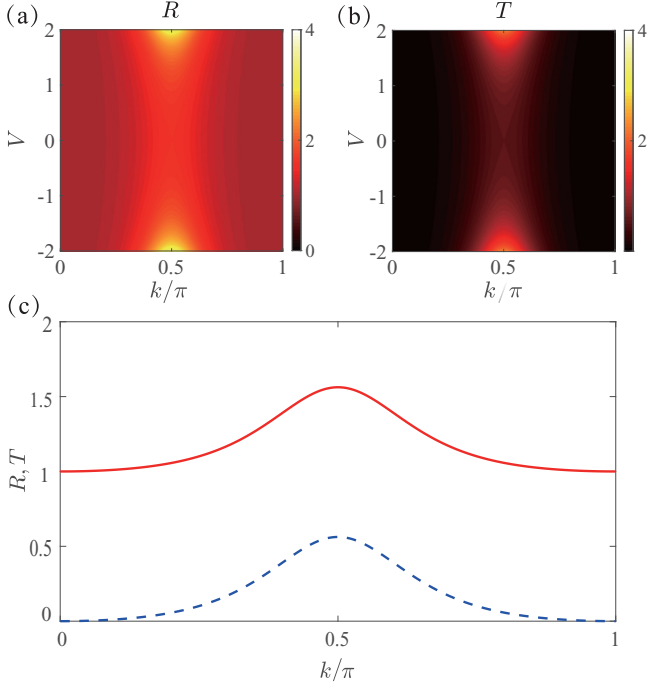

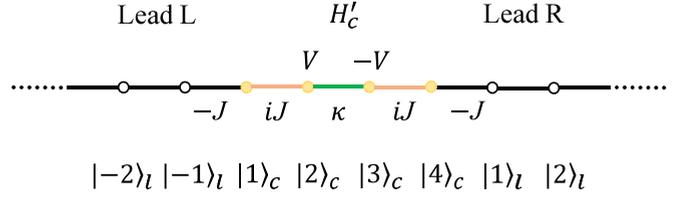

FIG. 3. Schematic of the anti-$\mathcal{PT}$-symmetric system with unitary scattering. The coupling sites between $|2\rangle_c$ and $|3\rangle_c$ is real $\kappa$ (green line). The scattering center $H'_c$ is still non-Hermitian.

FIG. 2. Plots for reflection and transmission in $H_c$. (a) and (b) are contour plots of the reflection $R(k, V)$ and the transmission $T(k, V)$ with $J = 1$, $\kappa = 3$ and $V \in [-2, 2]$. (c) is the reflection (solid red curve) and transmission (dashed blue curve) as a function of the input wave vector $k$ for $V = 0$.

$\psi_c^k(4) = e^{-ik} + r_R e^{ik}$. Substituting the wave functions into the Schrödinger equations, we obtain the reflection coefficients

$$r_L = \frac{-\kappa^2 + V^2 - 2iVJ\sin k - J^2 - 8J^2\cos^2 k}{4J^2\cos^2 k + J^2 e^{2ik} + \kappa^2 - V^2 + 4J^2\cos k e^{ik}}, \quad (13)$$

$$r_R = \frac{-\kappa^2 + V^2 + 2iVJ\sin k - J^2 - 8J^2\cos^2 k}{4J^2\cos^2 k + J^2 e^{2ik} + \kappa^2 - V^2 + 4J^2\cos k e^{ik}}, \quad (14)$$

and the transmission coefficients

$$t_L = t_R = \frac{2\kappa J \sin k}{4J^2\cos^2 k + J^2 e^{2ik} + \kappa^2 - V^2 + 4J^2\cos k e^{ik}}. \quad (15)$$

Notably, the reflection and transmission are both reciprocal, i.e., $t_L = t_R = t$, $|r_L| = |r_R| = |r|$. Furthermore, we obtain $|r|^2 - |t|^2 = R - T = 1$, which has a nonunitary scattering behavior and the excitation intensity is not conserved. We plot the reflection and transmission in Fig. 2 and we observe the conclusion $R - T = 1$.

## III. UNITARY SCATTERING

In the previous section, we elaborated on a four-site scattering center with imaginary couplings with anti-$\mathcal{PT}$-symmetry and nonunitary scattering behavior. In this section, we investigate an alternative anti-$\mathcal{PT}$-symmetric system, which is similar to the previous model. The only difference is that the coupling between the central two sites is replaced by the real coupling as shown in Fig. 3. In the following, we show the unitary scattering affected by the real coupling.

The Hamiltonian $H'$ of this model is

$$H' = H_{\text{lead}} + H_{\text{in}} + H'_c. \quad (16)$$

There is only a slight difference between the scattering centers $H_c$ and $H'_c$. In this model, the scattering center $H'_c$ reads

$$H'_c = iJ(|1\rangle_c\langle 2|_c + |2\rangle_c\langle 1|_c + |3\rangle_c\langle 4|_c + |4\rangle_c\langle 3|_c)$$
$$+ \kappa(|2\rangle_c\langle 3|_c + |3\rangle_c\langle 2|_c) + V(|2\rangle_c\langle 2|_c - |3\rangle_c\langle 3|_c) \quad (17)$$

and the matrix form of $H'_c$ is written as

$$H'_c = \begin{pmatrix} 0 & iJ & 0 & 0 \\ iJ & V & \kappa & 0 \\ 0 & \kappa & -V & iJ \\ 0 & 0 & iJ & 0 \end{pmatrix}. \quad (18)$$

The difference between two scattering centers $H_c$ and $H'_c$ is the coupling between sites $|2\rangle_c$ and $|3\rangle_c$. This coupling is a non-Hermitian imaginary coupling in $H_c$ and is a Hermitian real coupling in $H'_c$. Even though the coupling between sites $|2\rangle_c$ and $|3\rangle_c$ is changed, we notice that the scattering center $H'_c$ also possesses the anti-$\mathcal{PT}$ symmetry that $(\mathcal{P}'\mathcal{T})H'_c(\mathcal{P}'\mathcal{T})^{-1} = -H'_c$, where the parity operator $\mathcal{P}'$ is redefined as

$$\mathcal{P}' = \begin{pmatrix} 0 & 0 & 0 & 1 \\ 0 & 0 & 1 & 0 \\ 0 & -1 & 0 & 0 \\ -1 & 0 & 0 & 0 \end{pmatrix}. \quad (19)$$

Notably, $\mathcal{P}'$ is the generalized parity operator and the phase difference $e^{i\pi}$ exists between $|2\rangle_c$ and $|3\rangle_c$ (as well as $|1\rangle_c$ and $|4\rangle_c$) after the spatial inversion operation $\mathcal{P}'$.

To reveal the influence of central coupling, we calculate the reflection and transmission coefficients of $H'_c$. We still use the wave function in the form of Eqs. (7) and (8). The Schrödinger equations for the scattering center $H'_c$ are identical with $H_c$ in Eqs. (9) and (12), but Eqs. (10) and (11) change to

$$iJ\psi_c^k(1) + V\psi_c^k(2) + \kappa\psi_c^k(3) = E\psi_c^k(2), \quad (20)$$

$$\kappa\psi_c^k(2) - V\psi_c^k(3) + iJ\psi_c^k(4) = E\psi_c^k(3). \quad (21)$$

Then, we obtain the reflection coefficients

$$r_L = \frac{\kappa^2 + V^2 - 2iVJ\sin k - J^2 - 8J^2\cos^2 k}{4J^2\cos^2 k + J^2 e^{2ik} - \kappa^2 - V^2 + 4J^2\cos k e^{ik}}, \quad (22)$$

$$r_R = \frac{\kappa^2 + V^2 + 2iVJ\sin k - J^2 - 8J^2\cos^2 k}{4J^2\cos^2 k + J^2 e^{2ik} - \kappa^2 - V^2 + 4J^2\cos k e^{ik}}, \quad (23)$$

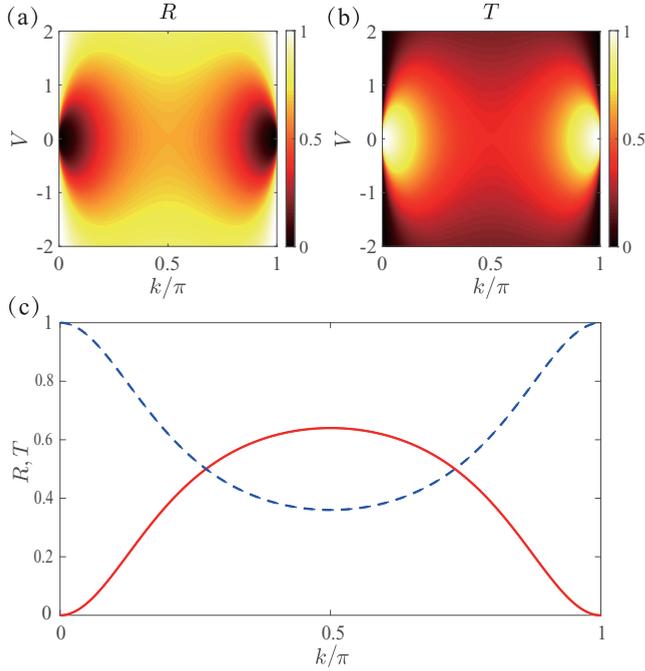

FIG. 4. Plots for reflection and transmission in $H'_c$. (a) and (b) are contour plots of the reflection $R(k, V)$ and the transmission $T(k, V)$ with $J = 1$, $\kappa = 3$ and $V \in [-2, 2]$. (c) is the reflection (solid red curve) and transmission (dashed blue curve) as a function of the input wave vector $k$ for $V = 0$.

and the transmission coefficients

$$t_L = t_R = \frac{-2i\kappa J \sin k}{4J^2 \cos^2 k + J^2 e^{2ik} - \kappa^2 - V^2 + 4J^2 \cos k e^{ik}}. \quad (24)$$

Notably, the reflection and transmission in this case are reciprocal, i.e., $t_L = t_R = t$, $|r_L| = |r_R| = |r|$. However, the reflection and transmission satisfy $R + T = 1$, which differs from the scattering coefficients for the scattering center $H_c$ in the previous section. The scattering dynamics exhibited in this case is unitary, being similar to the dynamics in a Hermitian scattering center [63]. The reflection and transmission of $H'_c$ are plotted in Fig. 4, the scattering is reciprocal and the excitation intensity is conserved.

Both two anti-$\mathcal{PT}$-symmetric scattering systems under consideration are transpose invariant and have time-reversal symmetry, which protect the symmetric transmission and symmetric reflection, respectively [62,64,65].

## IV. THE EFFECT OF COUPLING

From the scattering properties of the two above models, we notice that the coupling between the central two sites of the four-site scattering center plays a crucial role. First, the system is still anti-$\mathcal{PT}$ symmetric after altering the coupling between the central two sites. Then, for the non-Hermitian imaginary coupling of the structure in Fig. 1, it supports the nonunitary scattering with $R - T = 1$. If the non-Hermitian imaginary coupling $i\kappa$ between $|2\rangle_c$ and $|3\rangle_c$ alters the Hermitian real coupling $\kappa$ as illustrated in Fig. 3, the scattering center supports the unitary scattering with $R + T = 1$, similar to Hermitian systems even though the scattering center is still non-Hermitian. This indicates that non-Hermitian and Hermitian couplings between $|2\rangle_c$ and $|3\rangle_c$ significantly affect the scattering properties of anti-$\mathcal{PT}$-symmetric structures and induce the nonunitary and unitary scattering, respectively.

We emphasize that the essential point of the different scattering behaviors exhibited in the two models is the parity of the anti-$\mathcal{PT}$ symmetry, affected by the central couplings $i\kappa$ and $\kappa$. The two types of anti-$\mathcal{PT}$ symmetries are distinguished from the parities of the $\mathcal{PT}$ operators. From Eq. (6), we notice $(\mathcal{PT})^2 = I$, where $I$ is the identical matrix. Thus, the scattering center $H_c$ shown in Fig. 1 has even-parity anti-$\mathcal{PT}$ symmetry because of $(\mathcal{PT})H_c(\mathcal{PT})^{-1} = -H_c$. From Eq. (19), we note $(\mathcal{P}'\mathcal{T})^2 = -I$ and the scattering center $H'_c$ shown in Fig. 3 has odd-parity anti-$\mathcal{PT}$ symmetry because of $(\mathcal{P}'\mathcal{T})H'_c(\mathcal{P}'\mathcal{T})^{-1} = -H'_c$. The even-parity anti-$\mathcal{PT}$-symmetric scattering center has nonunitary scattering and the odd-parity anti-$\mathcal{PT}$-symmetric scattering center the unitary scattering.

Now we perform numerical simulations with a Gaussian wave packet injection. The initial excitation is a normalized Gaussian wave packet in the form of

$$|\phi(0, j)\rangle = \Omega_0^{-1/2} \sum_j e^{-(j-N_c)^2/2\sigma^2} e^{ik_c j}|j\rangle. \quad (25)$$

The Gaussian wave packet is centered at the site $N_c$, where $\Omega_0 = \sum_j e^{-(j-N_c)^2/\sigma^2}$ is the normalization factor, $k_c$ is the wave vector of the Gaussian wave packet, and the half width of the Gaussian wave packet is $2\sqrt{\ln 2}\sigma$ and characterizes its size. The time evolution of the Gaussian wave packet is

$$|\phi(t, j)\rangle = e^{-i\mathcal{H}t}|\phi(0, j)\rangle, \quad (26)$$

where $\mathcal{H}$ is Hamiltonian of a 100-site lattice including the four-site scattering center and two finite uniformly coupled leads connected to the scattering center. The wave propagating velocity $v = dE/dk$ is $2J \sin k_c$. As $H$ and $H'$ are both non-Hermitian, the evolution of Gaussian wave packet under $e^{-i\mathcal{H}t}$ is non-unitary.

In Fig. 5, a Gaussian wave packet is initially centered at the site $N_c = -25$. The wave vector for the Gaussian wave packet is $k_c = \pi/2$ and the velocity for the wave propagation is $v = 2J$. After the incident wave packet being scattered by the embedded scattering center around the site $j = 0$, it is split into a reflected wave and a transmitted wave that propagate in opposite directions. The intensity $|\phi(t, j)|^2$ of reflected and transmitted waves represent the reflection $R$ and the transmission $T$, respectively. For the scattering center $H_c$ shown in Fig. 5(a), the intensity difference between the reflected and transmitted waves is unity; for the scattering center $H'_c$ shown in Fig. 5(b), the intensity of the reflected and transmitted waves add up to unity. Therefore, the numerical simulations verify our analytical results $R - T = 1$ for the even-parity anti-$\mathcal{PT}$-symmetric scattering center and $R + T = 1$ for the odd-parity anti-$\mathcal{PT}$-symmetric scattering center.

In the following, we explore the dynamics of nonunitary scattering and show the dynamic features at the spectral singularity. Notably, the spectral singularity cannot exist in the unitary scattering, where the scattering exhibits Hermitian behavior. According to Eqs. (13)–(15), the divergence of $R$ and $T$ indicates the existence of the spectral singularity. We note that it occurs only at the point $\kappa^2 - V^2 = J^2$ for states with $k = \pm\pi/2$ and $\kappa/J > 1$ is the necessary condition for the existence of the spectral singularity. The corresponding wave functions are

$$\psi_{L,R}^{\pm\pi/2}(j) = \begin{cases} e^{i(\mp\pi/2)(j-1)}, & j < 0 \\ [(iV \mp J)/\kappa]e^{i(\pm\pi/2)(j+1)}, & j > 0. \end{cases} \quad (27)$$

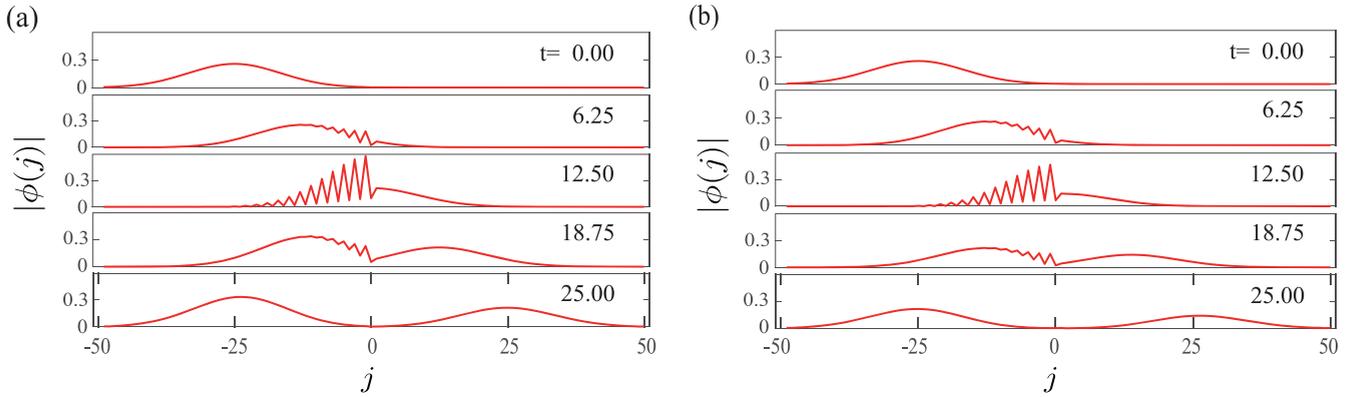

FIG. 5. The profile of time evolution for the injected Gaussian wave packet. (a) Scattering dynamics for $H_c$, the reflected intensity is $\sum_{j<0} |\phi(j)|^2 = 1.71$ and the transmitted intensity is $\sum_{j>0} |\phi(j)|^2 = 0.71$. (b) Scattering for $H'_c$, The reflected intensity is $\sum_{j<0} |\phi(j)|^2 = 0.70$ and the transmitted intensity is $\sum_{j>0} |\phi(j)|^2 = 0.30$. The parameters are $V = 1, \kappa = 3, J = 1$ for the system and $\sigma = 6, k_c = \pi/2$ for the wave packet. The relations $R - T = 1$ and $R + T = 1$ are verified in (a) and (b).

The physics of two states are clear, representing self-sustained emission and complete absorption of two oppositely propagating waves [23,66]. The intriguing feature of the spectral singularity is that two degenerate states for the incidences from the left and right merge into one state.

We perform the numerical simulations for the scattering dynamics at the spectral singularity. First, we consider a general solution for the wave injected in the left lead,

$$(e^{ik(j-1)} + re^{-ik(j-1)})|-j\rangle + te^{ik(j+1)}|j\rangle, \quad (28)$$

with $1 < j \leqslant N$. The reflection and transmission go to infinity for the momentum $k = \pi/2$. The solution corresponds to the wave emission dynamics of the initial state:

$$|\phi(0)\rangle = |\phi(N_c, \pi/2)\rangle. \quad (29)$$

The infinity of $r$ and $t$ should exhibit in the dynamics of the wave packet. Second, the solution $\psi_{L,R}^{-\pi/2}$ corresponds to the dynamics of two counter propagating wave packets initially centered at $\pm N_c$:

$$|\phi(0)\rangle = |\phi(N_c, \pi/2)\rangle + [(iV + J)/\kappa]|\phi(-N_c, -\pi/2)\rangle. \quad (30)$$

The profiles of the evolved states $|\phi(t)\rangle$ are plotted in Fig. 6.

In Fig. 6(a), an incident wave packet stimulates two counterpropagating emission waves with the amplitude ratio 1 : $(iV - J)/\kappa$. As $|(iV - J)/\kappa|^2 = 1$; the emission waves have the same amplitude. The intensities of reflected and transmitted waves increase linearly as time, which is a dynamical demonstration of the infinite reflection and transmission coefficients. Notably, the intensity difference between reflected and transmitted waves remains unity at the spectral singularity. In Fig. 6(b), a typical example of complete absorption is shown. Two incident waves with matching amplitudes and relative phases are fully absorbed after scattering.

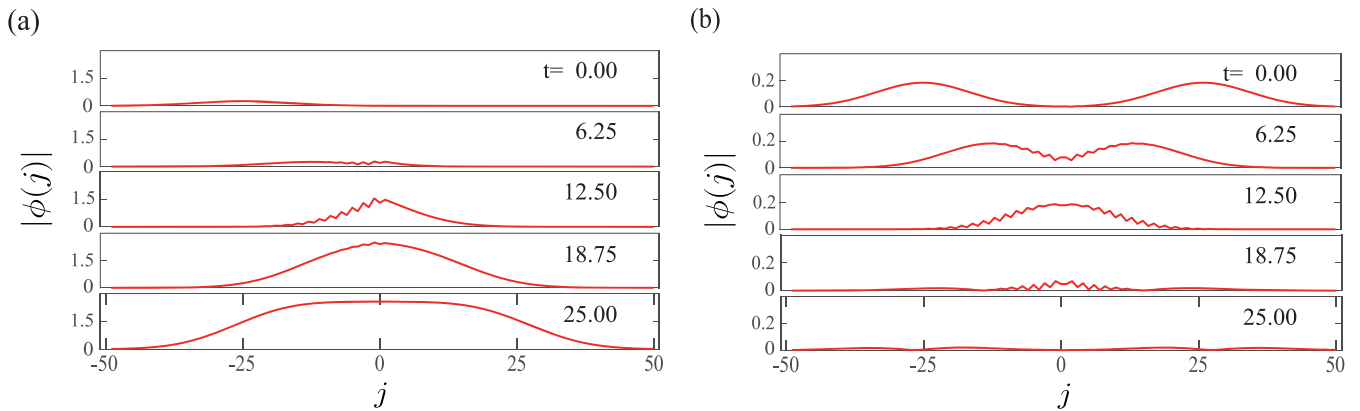

FIG. 6. The profiles of time evolution for the wave emission and perfect absorption. The time evolutions are governed by the Hamiltonian at the spectral singularity $J = 1, V = 1$, and $\kappa = \sqrt{2}$ for $k_c = \pi/2$. (a) The Gaussian wave packet in Eq. (29) with $N_c = -25$ and $\sigma = 6$. (b) Two Gaussian wave packets in Eq. (30) with the same parameter of $N_c$ and $\sigma$. In (a), the incident Gaussian wave packet is scattered at the center and forms a platform as the wave emission and $R - T = 1$ is always satisfied. In (b), the two Gaussian wave packets with a fixed phase difference are fully absorbed after collision.

## V. EXPERIMENTAL REALIZATION

The proposed anti-$\mathcal{PT}$-symmetric lattice models are implementable in many experimental platforms, including hybrid silicon microcavities [67–69], optical waveguides [70–72], and dielectric microwave resonators [73,74]. We focus on CROWs for the possible realization of the two anti-$\mathcal{PT}$-symmetric scattering systems.

CROWs are composed of primary resonators and linking resonators. All the resonators are evanescently coupled together and the primary resonators are indirectly coupled through the linking resonators. The primary resonators have an identical resonant frequency $\omega_0$ except for the resonators $|2\rangle_c$ and $|3\rangle_c$, which have frequencies $\omega_0 + V$ and $\omega_0 - V$, respectively. The dynamics in the proposed CROWs are described by the scattering Hamiltonians discussed in the previous sections as schematically illustrated in Figs. 1 and 3. Notably, each dot in the schematics represents a primary resonator and each line between the two neighbor primary resonators represents the effective coupling induced by the linking resonator after adiabatically eliminating the linking resonator. The effective coupling induced by the off-resonance passive linking resonator without dissipation/gain is real and Hermitian [67]; in contrast, the effective coupling induced by the on-resonance dissipative/active linking resonator with dissipative/gain is imaginary and non-Hermitian [56,60]. In the two situations, the effective coupling strength is approximately equal to the product of the two couplings between the linking resonator and its two adjacent primary resonators divided by the off-resonant frequency or the dissipation/gain rate of the linking resonator. The dynamics in the CROWs are, respectively, described by the Hamiltonians of the two anti-$\mathcal{PT}$-symmetric scattering systems.

## VI. CONCLUSION

We have investigated the scattering properties of two anti-$\mathcal{PT}$-symmetric non-Hermitian four-site scattering centers. The even-parity anti-$\mathcal{PT}$-symmetric scattering center possesses nonunitary scattering with $R - T = 1$; in contrast, the odd-parity anti-$\mathcal{PT}$-symmetric scattering center possesses unitary scattering with $R + T = 1$. The significantly different scattering dynamics is solely induced by the coupling between the central two sites of the scattering center, which determines the parity of the anti-$\mathcal{PT}$ symmetry. We emphasize that our conclusions are still valid for scattering centers with an even larger size as long as the scattering center structures $H_c$ and $H'_c$ keep unchanged with their difference being the couplings $i\kappa$ and $\kappa$ between the central two sites. Our findings deepen the understanding of anti-$\mathcal{PT}$ symmetry and its application in non-Hermitian physics, and the proposed systems can be verified in many experimental platforms including coupled waveguides, photonic crystals, and electronic circuits.

## ACKNOWLEDGMENT

This work was supported by the National Natural Science Foundation of China (Grant No. 11975128).